\documentclass[cup6a]{article}
\usepackage{graphicx}
\usepackage{natbib}
\catcode`\@=11
\def\gsim{\ifmmode{\,\mathrel{\mathpalette\@versim>\,}}
    \else{$\,\mathrel{\mathpalette\@versim>}\,$}\fi}
\def\lsim{\ifmmode{\,\mathrel{\mathpalette\@versim<\,}}
    \else{$\,\mathrel{\mathpalette\@versim<}\,$}\fi}
\def\@versim#1#2{\lower 2.9truept \vbox{\baselineskip 0pt \lineskip
    0.5truept \ialign{$\m@th#1\hfil##\hfil$\crcr#2\crcr\sim\crcr}}}
\catcode`\@=12

\def\LX{L_{\rm X}}

\title{Cusp-core dichotomy of elliptical galaxies: the role of thermal evaporation}
\author{Carlo Nipoti\\
{\small Dipartimento di Astronomia, Universit\`a di Bologna,}\\
\small{ via Ranzani 1, 40127 Bologna, Italy, E-mail: {\tt carlo.nipoti@unibo.it}}}

\begin{document}


\begin{center}
{\Large \bf Cusp-core dichotomy of elliptical galaxies: \\
the role of thermal evaporation}\\
\bigskip
{\large Carlo Nipoti\footnote{E-mail: {\tt carlo.nipoti@unibo.it}}}\\
\bigskip
{\small Dipartimento di Astronomia, Universit\`a di Bologna,}\\
\small{ via Ranzani 1, 40127 Bologna, Italy}
\end{center}
{\small May 28, 2009}


\begin{abstract}
There are two families of luminous elliptical galaxies: cusp galaxies,
with steep central surface-brightness profiles, and core galaxies,
whose surface-brightness profiles have flat central cores.  Thermal
evaporation of accreted cold gas by the hot interstellar medium may be
at the origin of this cusp-core dichotomy: in less massive (hot-gas
poor) galaxies central cores are likely to be refilled by central
starbursts following cold gas infall, while in more massive (hot-gas
rich) galaxies most cold gas is eliminated and central cores survive.
This scenario is consistent with the observation that cusp and core
galaxies differ systematically in terms of optical luminosity, X-ray
gas content, age of the central stellar population, and properties of
the active galactic nucleus.
\end{abstract}

\section{Introduction}

There are two families of luminous elliptical galaxies: cusp galaxies
and core galaxies. Cusp galaxies have steep power-law
surface-brightness profiles down to the centre (hence the name
``power-law'' galaxies often used as a synonymous of cusp galaxies),
corresponding to intrinsic stellar density profiles with inner
logarithmic slope $\gamma>0.5$; core galaxies have surface-brightness
profiles with a flat central core, corresponding to $\gamma<0.3$
\citep{Fab97,Lau07}.  Cusp galaxies are relatively faint in optical,
rotate rapidly, have disky isophotes, host radio-quiet active galactic
nuclei (AGN) and do not contain large amounts of X-ray-emitting gas;
core galaxies are brighter in optical, rotate slowly, have boxy
isophotes, radio-loud AGN and diffuse X-ray emission \citep[for a
  summary of these observational findings see][and references
  therein]{Nip07,Kor09}. The most popular explanation of the origin of
such a dichotomy is that cusp galaxies are produced in dissipative,
gas-rich (``wet'') mergers, while core galaxies in dissipationless,
gas-poor (``dry'') mergers \citep{Fab97}, the cores being a
consequence of core scouring by binary supermassive black
holes~\citep{Beg80}. The actual role of galaxy merging in the
formation of elliptical galaxies is still a matter of debate
\citep[e.g.][]{Naab09}. What is reasonably out of doubt is that cores
must be produced by dissipationless processes, while cusps are a
signature of dissipation~\citep[][and references
  therein]{Fab97,Kor09}. The dissipationless mechanism that forms the
cores must not necessarily be scouring by binary black holes: even
simple collisionless collapse may work \citep{Nip06}.  Dissipative
processes must be invoked to explain the observed dichotomy because
cusp galaxies are systematically less massive than core galaxies, and
no purely stellar-dynamical mechanism can introduce a characteristic
mass scale.  An interesting question is therefore why the dissipative
process responsible for the formation of cusps works in lower-mass
ellipticals, but does not work in the most massive
ellipticals. \citet{Nip07} argued that efficient thermal evaporation
of cold gas by the hot interstellar medium of the most massive
ellipticals can be at the origin of the cusp-core dichotomy.  Here I
briefly describe the basic principles of this scenario and discuss
some implications for the properties of AGN in elliptical galaxies.

\section{The formation of cusps and cores in elliptical galaxies}

\begin{figure}
\centering
\resizebox{1.\hsize}{!}{\includegraphics[clip=true]{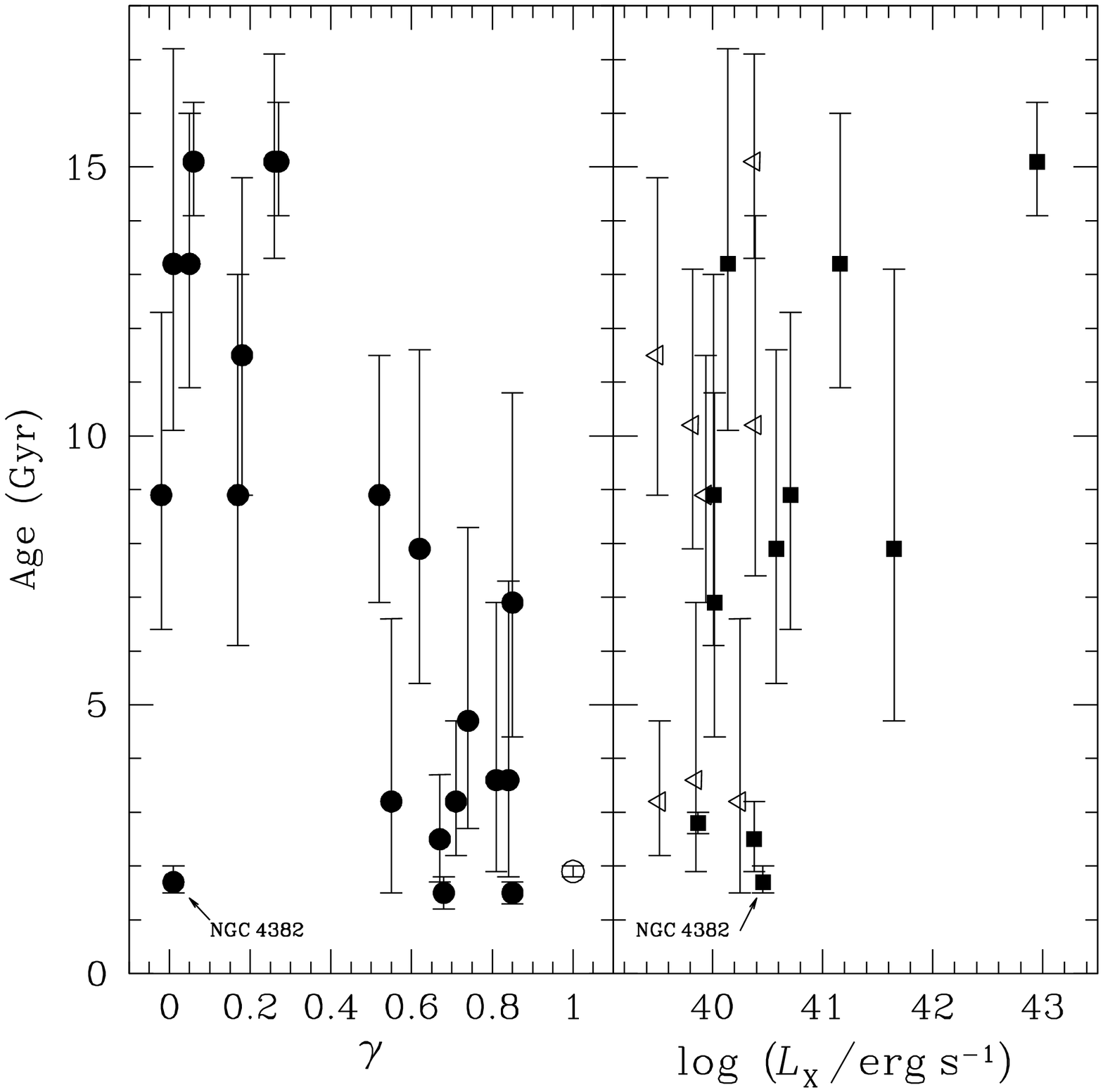}}
\caption{ The age of the central stellar population versus the central
  slope of the intrinsic stellar density profile $\gamma$
  \protect\citep[left-hand panel; adapted from][]{Nip07} and versus
  the soft X-ray luminosity $L_{\rm X}$ (right-hand panel) for
  early-type galaxies studied by \protect\cite{Mcd06}.  Central ages
  (with error bars) are from \protect\cite{Mcd06}, the values of
  $\gamma$ are from \protect\cite{Lau07} and the X-ray luminosities
  from \protect\cite{Pel05} and \protect\cite{Ell06}. For not all the
  galaxies measures of both $L_{\rm X}$ and $\gamma$ are available, so
  two slightly different sub-samples are represented in the left and
  right panels. In the right-hand panel solid squares represent
  detections and empty triangles upper limits in X-rays. The empty
  symbol in the left-hand panel represents NGC~4459, which is not in
  \protect\cite{Lau07} sample, but is classified as a cusp galaxy by
  \protect\cite{Kor09}: in the plot $\gamma=1$ is assumed arbitrarily
  just to indicate that it is a cusp galaxy.}
\label{fig:age}
\end{figure}

The stellar-dynamical process that forms the cores is expected to
operate at all mass scales: it is therefore natural to start from the
hypothesis that all ellipticals at some stage in their evolution have
central cores. Such cores can be later refilled by central bursts of
star formation, but a necessary condition for a central starburst to
happen is the availability of cold gas in the inner galactic
regions. Accretion of cold gas into galaxies, from minor mergers or
cosmic infall, is believed to be common even at late times. If nothing
prevents accreted cold gas reaching the galaxy centre, central
starbursts are likely and the core can be easily refilled by a
cusp. However, the journey of an accreted cold-gas cloud from the
outskirts down to the centre of an elliptical galaxy might be not so
safe. Massive elliptical galaxies are embedded in haloes of hot
(virial temperature) gas, and the interaction of cold ($T\lsim 10^4$
K) gas clouds with such a hot interstellar medium can disrupt the
clouds via a combination of ablation and evaporation by thermal
conduction.  The motion of a cold gas cloud through a hot plasma is a
complex dynamical process, involving heat conduction, radiative
cooling, ram-pressure drag and ablation through the Kelvin-Helmholtz
instability.  Less massive clouds are more vulnerable, and it is
possible to estimate with relatively simple analytic models a lower
limit to the minimum mass a cloud must have to survive evaporation,
depending on the temperature and density distribution of the hot
interstellar medium \citep{Nip07}.  The total mass of cold gas
available for central star formation is determined by the mass
spectrum of accreted gas clouds and by the minimum mass for survival
against evaporation.  The aggregate mass of new stars formed as a
consequence of cold infall is estimated to be proportionally larger in
lower-mass (hot-gas poor) ellipticals than in higher-mass (hot-gas
rich) ellipticals \citep{Nip07}. Thus cores are likely to be refilled
in the former, but not in the latter. This is consistent with the fact
that all galaxies with high X-ray emission from hot gas are core
galaxies, while ellipticals with lower X-ray emission include both
core and cusp galaxies \citep{Pel05,Ell06}.

If cusps are formed by late starbursts that refill a pre-existing
core, the central stellar population in cusp galaxies must be
relatively young.  This prediction is nicely verified in the sample of
early-type galaxies for which \cite{Mcd06} estimated the age of the
central stellar populations: in the plane of central age versus
central logarithmic slope $\gamma$ (Fig.~\ref{fig:age}, left-hand
panel) early-type galaxies from this sample are found to be neatly
segregated.  Core galaxies have median central ages 13.2 Gyr, while
cusp galaxies have median central age 3.6 Gyr \citep{Nip07}.  Given
the known anti-correlation between $\gamma$ and the soft X-ray
luminosity $\LX$ \citep[galaxies with high $\LX$ have
  $\gamma\lsim0.3$;][]{Pel05,Ell06}, it is interesting to see how
galaxies from the same sample are distributed in the plane of central
age versus $\LX$ (Fig.~\ref{fig:age}, right-hand panel): consistent
with the proposed scenario, there are no points in the bottom-right
area of the diagram. In other words, among the galaxies of this
sample, all those with young ($\lsim 5$ Gyr) central stellar
population have X-ray luminosity lower than $\sim 3 \times 10^{40}{\rm
  \,erg\,s^{-1}}$.

The only outlier of the bimodal distribution in the age-slope plane is
the lenticular galaxy NGC~4382, which is classified as a core galaxy,
but has young central stellar population (Fig.~\ref{fig:age},
left-hand panel).  It must be noted that NGC~4382 is quite peculiar
also in other respects: its very unusual morphology and surface
brightness profile suggest to interpret it as an unrelaxed recent
merger \citep{Kor09}.  Moreover, the diffuse X-ray luminosity of
NGC~4382 is relatively low \citep{Siv03,Pel05}, so the occurrence of a
central starburst is not surprising. One might speculate that in
NGC~4382 we are witnessing the first stages of the core-refill
process.

\section{Implications for active galactic nuclei in elliptical galaxies}

Thermal evaporation may also have a role in determining the mode of
accretion of central supermassive black holes in early-type galaxies.
It is widely accepted that there are two main modes of black-hole
accretion and feedback, usually referred to as ``cold mode'' (or QSO
mode) and ``hot mode'' \citep[or radio mode; e.g.][]{Bin05,Har07}. In
the cold mode the black hole feeds from cold gas, close to the
Eddington rate, and grows significantly in mass, with most of the
energy released going into photons (optical, UV, X-ray).  In the hot
mode the black hole feeds from hot gas, at a rate much below
Eddington's, and does not grow significantly in mass, with most of the
energy released being mechanical and generating significant radio
emission.  Bright QSOs and central radio sources in galaxy clusters
are prototypes of the cold and hot modes, respectively, but the basic
principles of this classification of the accretion modes apply to AGN
in general.

In the proposed picture cold gas can be available for accretion onto
the central black hole only in cusp galaxies. Thus, all core galaxies
must be hot-mode accretors, while cusp galaxies can be cold-mode
accretors, when there has been a recent episode of cold gas infall
into the galaxy.  This is consistent with the findings that the
optical nuclear emission (in units of the Eddington luminosity of the
central black hole) is typically higher by two orders of magnitude in
cusp galaxies than in core galaxies, that the nuclei of core galaxies
are radio-loud, while those of cusp galaxies are radio-quiet
\citep{Cap06,Balma06}, and that only core galaxies appear able to
produce powerful extended radio emission \citep{Der05}.  This model
fits in a more general scenario in which radio-loudness is more
controlled by the accretion mode than by black-hole spin, and hot-mode
AGN, similar to the micro-quasars observed in our galaxy, during their
lifetime alternate short bursts of radio-loudness with longer periods
of radio-quiescence \citep{Nip05,NipBB05}.

\section{Conclusions}

Since the discovery of the cusp-core dichotomy of elliptical galaxies
it has been proposed that cores are formed by dissipationless
processes, such as binary black hole scouring, while cusps are formed
by dissipative processes, such as merger-driven central starbursts
\citep{Fab97}. This proposal is consistent with several observed
properties of core and cusp galaxies, but does not explain {\it per
  se} why the most massive ellipticals are cored, while less massive
ellipticals are cusped. A possible explanation is that all ellipticals
originally have central cores: in less massive (hot-gas poor)
ellipticals the cores are refilled by central starbursts following
cold gas infall, while in more massive (hot-gas rich) ellipticals the
cores are preserved because the hot interstellar medium ablates and
evaporates most of the infalling cold gas \citep{Nip07}. In this
scenario black holes in core galaxies always accrete from hot gas,
while black holes in cusp galaxies can accrete from cold gas,
consistent with the observed properties of AGN in elliptical galaxies.
The importance of the hot gas in preserving the cores in the most
massive systems is supported by the state-of-the-art study of the
early-type galaxies in the Virgo cluster by \citet{Kor09}.

In the present paper attention has been focused on the the role of
thermal evaporation in determining the cusp-core dichotomy of
elliptical galaxies, but elimination of accreted cold gas by the hot
interstellar medium, via ablation and heat conduction, is likely to be
a fundamental process for galaxy formation in general, as it may be at
the origin of the truncation of the blue cloud and of the population
of the red sequence in color-magnitude diagrams of
galaxies~\citep{Bin04,Nip04,Nip07}. In this picture the role of black
holes in quenching star formation is fundamental, but indirect: via
AGN feedback they supply the hot gas with the energy necessary to
evaporate the cold gas and thus quench star formation.

\smallskip

I am grateful to James Binney for his comments on the manuscript and to 
Marc Sarzi for useful discussions.


\end{document}